\documentclass{caosp}
\usepackage{graphicx}

\def\teff{$T_{\rm eff}$}
\def\lgg{$\log g$}
\newcommand{\bs}{$\langle B_{\rm s}\rangle$}
\newcommand{\ms}{m\,s$^{-1}$}

\articleNo{123}
\pubyear{2005}
\volume{35}
\volnumber{3}
\firstpage{1}
\received{November 30, 2007}
\accepted{November 30, 2007}
 \def\ii{\,{\sc ii}} \def\iii{\,{\sc iii}}
\begin{document}
\hauthor{M.Sachkov {\it et al.}}

\title{Spectroscopic study of pulsations in the atmosphere of roAp star 10~Aql}

\author
{M.~Sachkov \inst{1}
\and O.~Kochukhov \inst{2}
\and T.~Ryabchikova \inst{1,3} \and
F.~Leone \inst{4}
\and S.~Bagnulo \inst{5}
\and W.W.~Weiss \inst{3}}
\institute
 {
Institute of Astronomy, Moscow, Russia, \email{msachkov@inasan.ru}
\and Department of Astronomy and Space Physics, Uppsala University, Sweden
\and Department of Astronomy, University of Vienna, Austria
\and Dipartimento di Fisica e Astronomia, Universit\`a di Catania, Italy
\and Armagh Observatory, Ireland}

\date{November 30, 2007}
\maketitle
\begin{abstract}
We present the analysis of spectroscopic time-series observations  of the roAp star 10
Aql. Observations were carried out in July 2006 with the UVES and SARG spectrographs simultaneously with
the MOST mini-satellite photometry.  All these data were analysed for radial velocity
(RV) variations. About 150 lines out of the 1000 measured reveal clear pulsation signal. Frequency analysis of the
spectroscopic data gives four frequencies. Three highest amplitude frequencies in
spectroscopy coincide with the photometric ones. Phase-amplitude diagrams created for
the lines of different elements/ions show that atmospheric pulsations may be
represented by a superposition of the standing and running wave components, similar to
other roAp stars. The highest RV amplitudes, 300--400~\ms, were measured for Ce\ii, Dy\iii, Tb\iii, 
and two unidentified lines at $\lambda\lambda$~5471, 5556 \AA.

We discovered $\approx$0.4 period phase jump in the RV measurements across the Nd\iii\
line profiles. It indicates the presence of the pulsation node in stellar
atmosphere. The phase jump occurs at nearly the same atmospheric layers for the two
main frequencies.

There is no rotational modulation in the average spectra for the 6 different nights we analysed.
\keywords{stars: atmospheres --
          stars: chemically peculiar --
      stars: magnetic fields --
      stars: oscillations}
\end{abstract}

\section{Introduction}
\label{intr}
10~Aql (HD~176232) was detected as rapidly oscillating Ap (roAp) star by Heller \& Kramer (1988) who
found three periods of $\approx$11.6, 12.1 and 13.4 min. Kochukhov et al. (2002) detected radial velocity (RV) pulsations with
amplitudes between 30 and 130~\ms\ and a period about 11.5 min. Later Hatzes \&  Mkrtichian (2005) confirmed
RV variations and registered the highest RV amplitude 398~\ms\ for an unidentified line at $\lambda$~5471.40~\AA. 
10 Aql was chosen for contemporaneous spectroscopic observations with large
ground based telescopes suited to obtain high time resolution,
high spectral resolution, and high signal-to-noise ratio spectra simultaneously
with high precision photometric observations with
MOST, the Canadian photometric space telescope (Walker et al., 2003).

The main photometric results
are published by Huber et al. (2007, these proceedings). Here we focus on the spectroscopic
analysis and are using MOST data primarily for a comparison of the frequency analysis results.

\section{Observations and data reduction}

Our observations of 10~Aql were obtained in 2006 with
the UVES spectrograph at the 8.2-m telescope UT2 (Kueyen),
of the VLT at Paranal (Chile) on July 3, 9, 15, and 17, and with the high resolution
spectrograph (SARG) at the 3.55-m {\it Telescopio Nazionale
Galileo} (TNG) at the Observatorio del Roque de los Muchachos
(La Palma, Spain) in July 14, 15, and 16.
For a frequency analysis we used an additional data set obtained also during the MOST
observing run with the UVES instrument on July 24 in the context of the observing programme 079.D-0567 
(ESO Archive was used to extract these data).

Each UVES data set but the last one consists of 211 spectra. The total number of spectra observed in July 24 is 105.
The peak signal-to-noise (S/N) ratio of individual spectra is up to 300, the
spectral region covered is 4960--6990~\AA\ (the
wavelength coverage is complete, except for a 100~\AA\, gap centred
at 6000~\AA), the resolving power is
$\lambda/\Delta\lambda\approx115\,000$, the time resolution (exposure and read out) is 70 s. 
We were able to measure radial velocity amplitude from a single line with the accuracy up to 10 \ms.
All UVES spectra were reduced and normalized to the continuum
level with a routine specially developed by D. Lyashko for a
fast reduction of time-series observations (Tsymbal et al., 2003). 
The total number of 207 spectra were obtained with SARG in the
4572--7922~\AA\ spectral range.  Resolving power is $\sim$100\,000, S/N is up to 150, 
and time resolution is 120 s. The
SARG spectra were reduced using standard
ESO-MIDAS software with the same main steps as described
above.

\section{Radial velocity measurements and frequency analysis}

\begin{figure}[t]
\centering{\includegraphics[width=8.5cm]{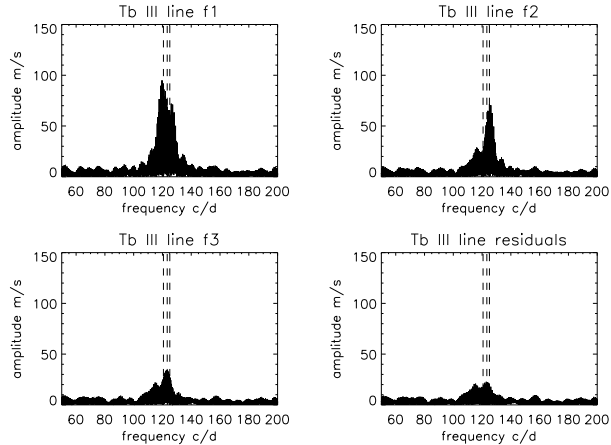}}
\caption{The amplitude spectra of RV variations of the Tb\iii\, 5505.370 \AA\, line. Dashed lines show position of the MOST photometric frequencies.} \label{dft}
\end{figure}

To perform a careful line identification and to choose lines for pulsation measurements we have synthesised 
the whole spectral region with the model atmosphere parameters 
\teff=7550 K, \lgg=4.0 and abundances from Ryabchikova et al. (2000) and magnetic field modulus \bs=1.5 kG 
(Kochukhov et al. 2002).

Both the centre-of-gravity and bisector methods were used for RV measurements. 
First, we measured practically all lines (about 2000) in July 03 observational set and detected about 150 lines that show pulsation signature. 
It is now well known that  pulsational variability is more pronounced in the lines
of rare-earth ions, especially those of Pr and Nd, which are strong
and numerous in the roAp spectra (see, for example, 
Kochukhov \& Ryabchikova, 2001). 10 Aql differs from others roAp stars in this context. REE lines are weak, 
REE abundances are lower than in other roAp stars, although the REE anomaly -- a characteristics of roAp stars, is present. 
Many of the pulsating lines have equivalent width less than 5 m\AA, therefore the errors of velocity measurements are rather high. 
For RV analysis we used the lines with equivalent width larger than 5 m\AA. The maximum RV amplitude, as large as 420~\ms, 
was detected in unidentified lines at $\lambda$~5471.41~\AA\ and $\lambda$~5556.13~\AA\ and in  Dy\iii~$\lambda$~5730.34~\AA\ line.
which all have equivalent widths around 9--11 m\AA\ and residual depth around 6--7\%. 

Due to similarity of
the pulsation patterns in the unidentified and Dy\iii\ lines, one may suppose that they should belong to the same atomic species. 
Strong Nd\iii\, and Pr\iii\ lines, that usually exhibit largest amplitudes in roAp stars show lowest amplitudes in 10 Aql
(probably due the existence of nodal zone, as will be shown below). Nd\ii\ and Pr\ii\ lines are very weak. Hence, the blending problem in 
this star becomes more acute. Even slight blending results in abrupt decrease of the amplitude. Therefore, for the final amplitude and phase 
analysis only 70 cleanest lines were chosen. 

We performed frequency analysis of our measurements by applying the
standard discrete Fourier transformation (DFT) to the RV data. The total time 
coverage of the spectroscopic data is about 25 hours. The
period corresponding to the highest pulsation amplitude was
improved by the sine-wave least-square fitting of the RV data
with pulsation period, amplitude, and phase treated as free
parameters. This fit was removed from the data and then Fourier
analysis was applied to the residuals. This procedure was repeated
for all frequencies with the S/N above 5 in the power spectrum. The primary frequency (f$_1$) was found at 119.69 c/d, which
corresponds to a 1-day alias of the frequency seen in simultaneous MOST photometry. 
The second signal (f$_2$) is at 125.09 c/d and the third frequency (f$_3$) is at 123.30 c/d. 
All three frequencies appear in RV data for most pulsating lines. No clear signal can be found in the data 
after pre-whitening of the three frequencies, although in  the residuals for several lines  some signal appears
near 117--118 c/d and 126--127 c/d. But this signal has S/N$<$2. 
Figure\,\ref{dft} shows an amplitude spectrum for the Tb\iii\ 5505.370 \AA\ line; 
the other panels show the next prewhitening steps.

The MOST photometric observations give the same frequencies. The only difference is that the
frequency f$_2$ has the highest amplitude in the MOST data and f$_1$ is the second one (Huber et al. 2007). This can 
be explained by the different time coverage of the spectroscopic data compared to continuous photometric set. Making a frequency analysis
of the part of photometric data taken at the times of spectroscopic monitoring 
one obtains f$_1$ with the highest amplitude in photometric data, too (Huber, private communication). 

\begin{figure}[!t]
\hbox{\vspace{-2mm}
\centering{\includegraphics[width=8cm]{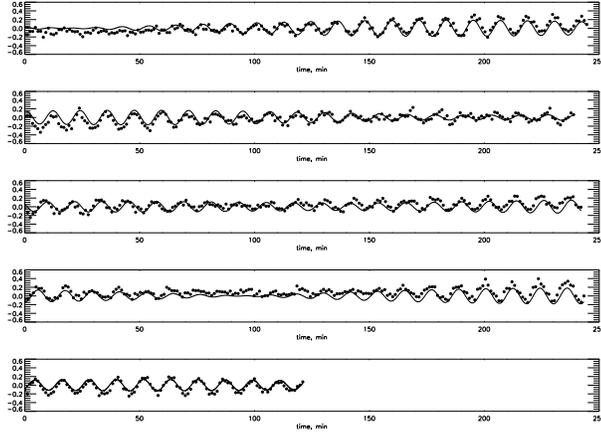}}
\hfill
\parbox[b]{32mm}{
\caption{The RV variations of the Tb\iii\ 5505.370 \AA\ line.} \label{beat}}}
\end{figure}

We have to stress the importance of simultaneous precise photometry for the spectroscopic time-series study, in particular 
to avoid alias problems in frequency analysis. 
Moreover, one relatively short spectroscopic set, that is usually no longer than 2--4 hours, does not allow one to resolve 
close frequencies. Wrong conclusions can be made from amplitude modulations that caused just by the beating effect. 
In the case of 10~Aql, a superposition of the three close frequencies fully explains the observed amplitude modulation, which is
illustrated in  Fig.\,\ref{beat} where RV variations 
of the Tb\iii~5505.370 \AA\ line are presented for UVES data (from top to bottom for nights July 3, 9, 15, 17, 24; 
the reference time point for each data set is the HJD time of the first observation). 
The solid line  shows sinusoidal fit including all three frequencies.

\begin{figure}[!th]
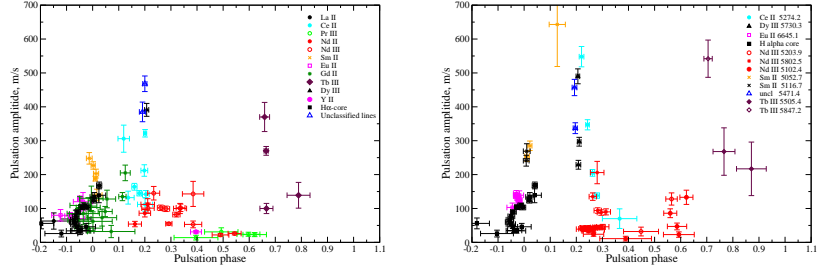

\centerline{\includegraphics[width=5cm,clip=]{jul03_rv-ph_per1.eps}
\hspace{0.5cm}
\includegraphics[width=5cm,clip=]{jul03_bis.eps}}
\caption{Amplitude-phase diagram for the f$_1$ frequency in the July 03 UVES data. Left panel: centre-of-gravity measurements,
right panel: bisector measurements.}
\label{jul03}
\end{figure}

\section{Pulsational analysis}

We did not found any evidence of the rotational modulation during a month of our spectroscopic observations. The average spectra in all sets 
agree within better than 0.5\% which means that rotation period should be an order of several months at least.

Following Ryabchikova et al. (2007a), we analysed
the amplitude-phase diagrams and interpreted observations in
terms of pulsational wave propagation. The pulsational behaviour of 10~Aql is quite similar to the one found in other roAp stars: pulsation 
appears in the layers where La and Eu are concentrated, then goes through the layers where H$\alpha$-core is formed, reaches maximum of RV
and then the amplitude decreases (see Ryabchikova et al. 2007b). At the same time, the RV maximum is attained in Ce\ii\, and Dy\iii\, lines and not
in Nd\ii\, Nd\iii\, Pr\iii\, lines, as it is observed in most other roAp stars.  In the layers where Nd and Pr lines are formed pulsation 
amplitude falls practically to zero, and this is accompanied by the rapid phase change. When RV amplitude increases again in Tb\iii\ lines, the
phase changes by $\sim$0.4 (left panel in Fig.\,\ref{jul03}). We attribute this phenomenon to the presence of a node. Similar amplitude-phase diagrams were
obtained for the f$_2$ frequency. The phase-amplitude diagrams are also similar for all 4 nights of our UVES observations. 

\begin{figure}
\centerline{\includegraphics[width=9cm]{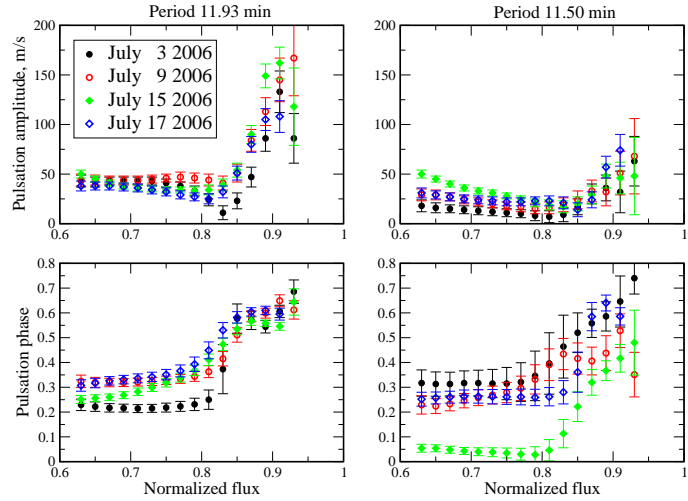}}
\caption{Bisector measurements for the Nd\iii~5102.435 \AA\ line.}
\label{bis_Nd}
\end{figure}

The same picture was derived from the bisector measurements of individual lines (right panel in Fig.\,\ref{jul03}). 
As it was mentioned earlier, the REE lines
usually showing the highest pulsation signal are weak in 10~Aql. Fortunately, there is sufficiently deep line, Nd\iii~5102.435 \AA,
suitable for precise bisector measurements. Although the centre-of-gravity RV amplitude is very small (30--40~\ms), the bisector amplitude changes across the line profile
with the minimum around 0.83 of the normalized flux. RV amplitude change is accompanied by the phase jump. We found the same behaviour in all 4 nights 
and in both of the highest highest amplitude frequencies f$_1$ and f$_2$. Fig.\,\ref{bis_Nd} shows bisector measurements for the Nd\iii~5102.435 \AA\ line. 
Thus, 10~Aql is the second roAp star after 33~Lib (Mkrtichian et al. 2003) which shows direct evidence of an atmospheric node.

\acknowledgements
This work was supported by the
 RFBI (grant 06-02-16110a), by the
Swedish \textit{Kungliga Fysiografiska S\"allskapet}, by
\textit{Royal Academy of Sciences} (grant No. 11630102), and by
Austrian Science Fund (FWF-P17580).

\end{document}